\documentstyle[12pt]{article}

\begin{document}


\begin{center}
{\Large {\bf {Four Dimensionality in Non-compact Kaluza-Klein Model}}}\\
\vskip 0.5cm {\ Merab GOGBERASHVILI } \\
\vskip 0.3cm {\it {Institute of Physics, Georgian Academy of Sciences} \\
{6 Tamarashvili str., Tbilisi 380077, Georgia}\\
{(E-mail: gogber@hotmail.com)} } \\
\vskip 0.5cm

{\Large {\bf Abstract}}\\
\vskip 0.3cm

\quotation
{\small Five dimensional model with extended dimensions investigated. It is
shown that four dimensionality of our world is the result of stability
requirement. Extra component of Einstein equations giving trapping solution
for matter fields coincides with the one of conditions of stability. } 
\endquotation

\end{center}

\vskip 0.5cm 

\section{Introduction}

An investigation of possibility that dimension of our world is more then
four is not new. Nearly all papers on this direction are done with the
framework of standard Kaluza-Klein models where extra dimensions are curled
up to an unobservable size (see for example review \cite{OW}). Besides of
obvious achievements this approach encounters some problems such as: Why
four dimensions are extended and others are curled; How to choose the
signature of multidimensional space; Physical meaning of extra components of
Einstein's equations is unclear; There exists the problem of stability.

An alternative proposal that the extra dimensions are extended and the
matter is trapped in 4-dimensional submanyfold was advanced in papers 
\cite{RS,V,S,BK}. This approach has properties similar to four dimensions - all
dimensions are extended and equal at the beginning and the signature has the
form (+,-,-, ... ,-).

Models of this kind also do not contradict to present time experiments 
\cite{OW}. Multidimensionality in these models was used to solve several
problems, such as, cosmological constant, dark matter, non-locality or
hierarchy problems \cite{RS,V,S,KH,G}.

For the simplicity, we investigate here only the case of five dimensions.
The general procedure immediately generalizes to arbitrary dimensionality.

Using extended dimensions approach we want to consider our Universe as a
3-shell expanding in 5-dimensional space-time \cite{G}. This model supported
by at least two observed facts. First is the isotropic runaway of galaxies,
which only for obviousness usually explained as an expansion of a 3-sphere
in five dimensions. Second is the existence of a preferred frame in the 
Universe where the relict radiation is isotropic. In the framework of the
closed-Universe model without boundaries this can also be explained if
Universe is a bubble and the mean velocity of the background radiation is
zero with respect to its center in the fifth dimension.


\section{Stability Condition}

\setcounter{equation}{0}

In models of large extra dimensions we need the mechanism of confining a
matter inside of the 4-dimensional submanyfold which must be sufficiently
narrow along the extra dimensions and flat along four others. It is natural
to think that such a splitting of 5-dimensional space with trapping of a
matter into four dimensions is the result of existence of the special
solution of multidimensional Einstein equations 
\begin{equation}  \label{1.1}
^5R_{AB}-\frac{1}{2}g_{AB}~^5R = 6\pi ^2GT_{AB}~~.
\end{equation}
Here $G$ is 5-dimensional gravitational constant and $A,B,...=0,1,2,3,5$.

We need stabile macroscopic solution, so it is natural to consider only
classical gravitational, electromagnetic and scalar fields which can form
extended solutions. For the beginning let us consider only gravitational and
electromagnetic fields with the Lagrangian 
\begin{equation}  \label{1.3}
L= - \sqrt{g}\left[ \frac{1}{12\pi ^2G}~^5R + 
\frac{1}{4}F_{AB}F^{AB}\right] ~~.
\end{equation}
Generallisation for any Yang-Mills field is obvios.

To obtain stabile splitting of multidimensional space momentum toward the
extra - fifth dimensions must be zero 
\begin{equation}  \label{1.4}
P_5=\int T_5^5dtdV+\int T_5^\alpha dS_\alpha =0
\end{equation}
(Greek indices $\alpha ,\beta ...=0,1,2,3$ numerate coordinates in four
dimensions). Other dimensions can expand and our world can be expanding
bubble.

For our choice of gravitational Lagrangian the energy-momentum tensor of
gravitation and electromagnetic fields in five dimensions $T_{AB}$ has the
form of so named Lorentz energy-momentum complex 
\begin{equation}  \label{1.5}
T_A^B=t_A^B+\tau_A^B=\frac{1}{12\pi^2G\sqrt{g}}\partial_CX_A^{BC}~~,
\end{equation}
where 
\begin{equation}  \label{1.6}
X_A^{BC}=-X_A^{CB}=\sqrt{g}[g^{BD}g^{CE}(\partial _Dg_{AE}-
\partial _Eg_{AD})]~~.
\end{equation}

In (\ref{1.5}) 
\begin{eqnarray}
t_{A}^{B} = \frac{1}{12\pi^2G}(g^{BD}\partial_{A}\Gamma^{E}_{DE} -
g^{ED}\partial_{A}\Gamma^{B}_{DE} + \delta^{B}_{A}~^5R)~~,  \nonumber \\
\tau_{A}^{B} = - F^{BC}F_{AC} + \frac{1}{4}\delta^{B}_{A} F^{DE}F_{DE}
\label{1.7}
\end{eqnarray}
are energy-momentum tensor of gravitational and electromagnetic fields
respectively.

To obey the stability condition (\ref{1.4}) for the solutions we must have 
\begin{equation}  \label{1.8}
T_5^\alpha =T_5^5=0~~.
\end{equation}
Using (\ref{1.6}) from (\ref{1.8}) we obtain 
\begin{equation}  \label{1.9}
\partial_\beta g_{5A}=\partial_5g_{5\alpha }=0~~.
\end{equation}
Simple solution of (\ref{1.9}) is 
\begin{equation}  \label{1.10}
g_{5\alpha }=0~~,~~~g_{55}=const=-1~~.
\end{equation}
In general $g_{55}$ can be any function of $x^5$, but this function in all
formulae will appear as a coefficient and would not influence on the
dynamic of the theory.

From the other hand from first condition of (\ref{1.8}) using explicit form 
(\ref{1.7}) we obtain 
\begin{equation}  \label{1.11}
\partial_5\Gamma_{\beta \gamma }^\alpha =0~~,~~~F_{5\alpha }=0
\end{equation}
and equation of electromagnetic field has standard Maxwell 4-dimensional
form 
\begin{equation}  \label{1.12}
D_\nu F^{\mu \nu }=0~~.
\end{equation}

Here we want to notice that in model of Visser \cite{V} electromagnetic
field don't obey the condition (\ref{1.11}) ($F_{5\alpha }$ not equal to
zero) and therefore his solution is unstable.

Finally from (\ref{1.10}) and first of (\ref{1.11}) we obtain the metric
tensor corresponding to stable splitting of multidimensional space-time 
\begin{equation}  \label{1.13}
g_{\alpha \beta }=\lambda ^2(x^5)\eta_{\alpha \beta }~~,
~~~g_{55}=-1~~,~~~g_{5\beta }=0~~,
\end{equation}
where $\eta_{\alpha \beta }$ is the 4-dimensional metric tensor and 
$\lambda ^2(x^5)$ in the meanwhile is the arbitrary function of fifth
coordinate. This solution which we received from stability conditions
exactly coincide with the anzats of Rubakov-Shaposhnikov \cite{RS}.

Splitting (\ref{1.13}) was made in the frame of the 4-dimensional wall. If
we consider our Universe as an expanding bubble in the frame of the bubble
center (\ref{1.13}) needs Lorentz transformation. Bubble expansion $\lambda
(x^5)\rightarrow \lambda (x^5-vt)$, where $v$ is velocity of the wall, means
conformal transformation of 4-dimensional metric $\eta_{\alpha \beta }$. To
keep stability, the theory must be invariant under conformal transformations
in the submanyfold. This condition fixes dimension of our world. Indeed
using formulae (\ref{1.12}) and (\ref{1.13}) Lagrangian of electromagnetic
field in any dimensions can be written in the form 
\begin{equation}  \label{1.14}
L=\sqrt{\lambda ^{2n}\eta }\frac {1}{4\lambda ^4}\eta^{\alpha 
\gamma }\eta^{\beta \delta }F_{\alpha \beta }F_{\gamma \delta }~~,
\end{equation}
where $n$ is dimension of submanyfold of trapping. Only in case of four
dimensions ($n=4)$ we have conformal invariance and stabile splitting is
possible. Thus only 3-dimensional expanding bubbles can be survived for a
long time and our Universe can be one of them.


\section{Trapping}

\setcounter{equation}{0}

In previous section it was shown that in Gausian normal coordinates the
5-dimensional metric of our Universe can be written in the form 
\begin{equation}  \label{2.1}
ds^2=-(dx^5)^2+\lambda ^2(x^5) \eta_{\alpha \beta }dx^\alpha dx^\beta ~~.
\end{equation}
In this coordinates components of Christoffel symbols with two or three
indices $5$ are equal to zero, while with the one index $5$ forms the tensor
of extrinsic curvature \cite{MWT} 
\begin{eqnarray} 
\nonumber K_{\alpha\beta} = \Gamma^{5}_{\alpha\beta } = 
\frac{1}{2}\partial_{5}g_{\alpha\beta } = 
\lambda\lambda^{^{\prime}}\eta_{\alpha\beta }~~, \\
\label{2.2}K^{\alpha\beta} = - \frac{1}{2}\partial_{5}g^{\alpha\beta } ~~,
\end{eqnarray}
where prime denotes derivative with respect of the coordinate $x^5$. Also we
want to represent some useful formulae 
\begin{eqnarray}  \label{2.3}
g^{\alpha\gamma}K_{\gamma\beta} = \Gamma^{\alpha}_{5\beta } =
\lambda\lambda^{^{\prime}}\delta^{\alpha}_{\beta }~~,  \nonumber \\
K = g^{\alpha\beta}K_{\alpha\beta} = g_{\alpha\beta}K^{\alpha\beta} =
4\lambda^{^{\prime}}/\lambda~~, \\
\partial_5 K= g^{\alpha\beta}\partial_5 K_{\alpha\beta} -
2K^{\alpha\beta}K_{\alpha\beta}~~.  \nonumber
\end{eqnarray}

Any vector and tensor naturally is split-up into its components orthogonal
and tangential to the shell. Using decomposition of the curvature tensor 
\begin{eqnarray}  \label{2.4}
^5R_{\alpha\beta} = R_{\alpha\beta } + \partial_5 K_{\alpha\beta } -
2K_{\alpha}^{\gamma} K_{\gamma\beta } + K K_{\alpha\beta }~~,  \nonumber \\
^5R_{55} = - \partial_5 K - K^{\alpha\beta}K_{\alpha\beta}~~, \\
^5R = R + K^{\alpha\beta}K_{\alpha\beta} + K^2 + 2\partial_5 K  \nonumber
\end{eqnarray}
one can find decomposition of Einstein's equations 
\begin{eqnarray}  \label{2.5}
R_{\alpha\beta } - \frac{1}{2}\eta_{\alpha\beta}R + \partial_5 K_{\alpha\beta }
- 2K_{\alpha}^{\gamma} K_{\gamma\beta } + K K_{\alpha\beta } - \frac{
\lambda^2}{2}\eta_{\alpha\beta }( K^{\gamma\delta}K_{\gamma\delta} + K^2 +
2\partial_5 K) =  \nonumber \\
= \frac{6\pi ^2G}{\lambda^2}(- F^{\delta}_{\alpha}F_{\delta\beta} + 
\frac{1}{4}\eta_{\alpha \beta } F^{\gamma\delta}F_{\gamma\delta} )~~ , \\
\frac{1}{2\lambda^2}R + \frac{1}{2}(K^2 - K^{\gamma\delta}K_{\gamma\delta})
= - \frac{3\pi ^2G }{2\lambda^4} F^{\alpha\beta }F_{\alpha\beta }~~ . 
\nonumber
\end{eqnarray}

Using last formula of (\ref{2.3}) we noticed that since
\begin{equation}
t_{5}^{5} = \frac{1}{12\pi ^2G}(- \partial_{5}K -
g^{\alpha\beta}\partial_{5}K_{\alpha\beta} + ~^5R)~~, 
\label{2.51}
\end{equation}
fifteenth Einstein's equation (last equation of (\ref{2.5})) is nothing else 
than stability condition (\ref{1.8}) - $T_5^5=0$. So extra
component of Einstein's equations, whose physical meaning is unclear in
standard Kaluza-Klein models, in our approach coincide with the condition of
stability.

Using the explicit form of extrinsic curvature tensor (\ref{2.3}) from the
system of Einstein's equations (\ref{2.5}) we find 
\begin{equation}
-R=12\lambda \lambda ^{"} ~~.  \label{2.6}
\end{equation}
Then from second equation of (\ref{2.5}) for function $\lambda (x^{5})$ we
have 
\begin{equation}
\lambda \lambda ^{"}- \lambda ^{^{\prime }2}-\frac{\pi ^{2}G}{4}F^{\gamma
\delta }F_{\gamma \delta } =0~~.  \label{2.7}
\end{equation}

This equation gives trapping solution 
\begin{equation}  \label{2.8}
\lambda =cosh(Ex^5)~~,
\end{equation}
where 
\begin{equation}  \label{2.9}
E=\sqrt{\frac{\pi ^2G}4F^{\gamma \delta }F_{\gamma \delta }}~~.
\end{equation}

Width of our world $\Delta \sim 1/E$ depended on gravitational constant and
density of electromagnetic field.

To see how gravitational trapping works let us consider simple example of
the real scalar field $\phi $ in the background metric (\ref{1.13}) 
with $\lambda $ expressed by (\ref{2.8}). If we put 
\begin{equation}  \label{2.10}
\phi =\lambda u(x^\alpha )
\end{equation}
to the 5-dimensional Klein-Gordon equation 
\begin{equation}  \label{2.11}
\frac 1{\sqrt{g}}\partial _A(\sqrt{g}g^{AB}\partial _B\phi )+m^2\phi =0~~,
\end{equation}
for $u$ we find 
\begin{equation}  \label{2.12}
\eta^{\alpha \beta }\partial _\alpha \partial _\beta u+
\frac{m^2\lambda ^4-1}{\lambda ^2}u=0~~.
\end{equation}
According to (\ref{2.8}) - $\lambda =cosh(Ex^5)$ and we see that ''mass'' of
the scalar field in this equation has its minimum in the 4-dimensional
submanyfold and growth rapidly far from the wall. So field $\phi $ is in the
potential well.


\section{Scalar fields}

\setcounter{equation}{0}

Now let us consider more complicate model adding to (\ref{1.3}) the
Lagrangian of complex scalar fields 
\begin{equation}  \label{3.1}
L_\psi =\sqrt{g}[D_A\bar \psi D^A\psi -\xi |\psi |^2~^5R-U(|\psi |)]~~,
\end{equation}
where $U(|\psi |)$ in the meanwhile is any function of $\bar \psi \psi $.
Variation of this Lagrangian by metric tensor gives energy-momentum tensor
of scalar fields 
\begin{equation}  \label{3.2}
T_{AB}=t_{AB}+2\xi (~^5R_{AB}-\frac 12g_{AB}~^5R+D_AD_B-g_{AB}D_CD^C)|\psi |^2~~,
\end{equation}
where 
\begin{equation}  \label{3.3}
t_{AB}=D_A\bar \psi D_B\psi +D_B\bar \psi D_A\psi -g_{AB}[D_C\bar \psi
D^C\psi -U(|\psi |)]~~.
\end{equation}

Stability condition for the shell (\ref{1.8}) - $T_{5\alpha }=0$, now except
of conditions $~^5R_{5\alpha }=F_{5\alpha }=0$ gives the new condition for
scalar fields 
\begin{equation}  \label{3.4}
D_5\psi =(\partial _5-iA_5)\psi =0~~.
\end{equation}

Maxwell equations now take the standard 4-dimensional form with the source 
\begin{equation}  \label{3.5}
D_\mu F^{\mu\nu }=j^\nu =\bar \psi \partial ^\nu \psi -\psi \partial ^\nu
\bar \psi ~~,
\end{equation}
and the equation for the scalar field is 
\begin{equation}  \label{3.6}
(\eta^{\mu \nu }\partial _\mu \partial _\nu +\xi ~^5R)\psi +
\lambda ^2\frac{\partial U(|\psi |)}{\partial \bar \psi }~=0~~.
\end{equation}

Using formulae (\ref{2.3}) for the extrinsic curvature tensor $K_{\mu \nu }$, 
splitting of Einstein's equations (\ref{2.5}) now has the form 
\begin{eqnarray}  
\left( \frac{1}{6\pi ^2G } + 2\xi|\psi|^2\right)(R_{\alpha\beta } - 
\frac{1}{2}\eta_{\alpha \beta }R) + 
\frac{1}{2\pi ^2G } \eta_{\alpha \beta } \frac{\lambda^{"}}{\lambda} =  \nonumber \\
= \frac{1}{\lambda^2}(- F^{\delta}_{\alpha}F_{\delta\beta} + 
\frac{1}{4} \eta_{\alpha\beta} F^{\gamma\delta }F_{\gamma\delta} ) + 
D_{\alpha}\bar{\psi}D_{\beta}\psi + D_{\beta}\bar{\psi}D_{\alpha}\psi -  \nonumber \\
\label{3.7}- \eta_{\alpha\beta}[\eta^{\mu\nu}D_{\mu}\bar{\psi}D_{\nu}\psi -
\lambda^2U(|\psi|)] + 2\xi(D_{\alpha}D_{\beta} -
\eta_{\alpha\beta}D_{\mu}D^{\mu}) |\psi|^2 ~~, \\
\left( \frac{1}{6\pi ^2G } + 2\xi|\psi|^2\right)\frac{1}{2\lambda^2}R + 
\frac{1}{\pi ^2G }\frac{(\lambda^{^{\prime}})^2}{\lambda^2} =  \nonumber \\
\label{3.8}= - \frac{1}{4\lambda^4} F^{\alpha\beta }F_{\alpha\beta } + 
\frac{1}{\lambda^2}\eta ^{\mu\nu}D_{\mu}\bar{\psi}D_{\nu}\psi - U(|\psi|) - 
\frac{2\xi}{\lambda^2}D_{\mu}D^{\mu}|\psi|^2 ~~.
\end{eqnarray}

Substituting of (\ref{3.6}) and (\ref{3.8}) in trace of equation (\ref{3.7})
one can find that solution of the system (\ref{3.6}) - (\ref{3.8}) is 
\begin{eqnarray}  \label{3.9}
6\xi = 1~~,  \nonumber \\
R = -12\lambda \lambda^{"} ~~, \\
\lambda = cosh(Ex^5)~~ .  \nonumber
\end{eqnarray}

We see that constant of coupling gravitational and scalar fields $\xi $
fixed on the value corresponding to conformal invariance of scalar field
equation in four dimensions. So again we received 4-dimensionality from the
condition of stability towards the extra dimensions. Only conformal
invariant form of function $U(|\psi |)$ in four dimensions, when $\psi
=\lambda (x^5)/u(x^\nu )$, is 
\begin{equation}  \label{3.10}
U(|\psi |)=\mu|\psi |^4/2~~,
\end{equation}
where $\mu$ is coupling constant. Finally equation of massless scalar field
in five dimensions (\ref{3.6}) has the form 
\begin{equation}  \label{3.11}
(\eta ^{\mu \nu }D_\mu D_\nu +2E^2+\mu |\psi |^2)\psi =0~~.
\end{equation}
We see that because of coupling with gravitational field in four dimensions
scalar field has "mass" $E^2$ expressed with gravitational constant and
density of electromagnetic field by (\ref{2.9}).


\end{document}